\documentclass[10pt,letterpaper]{article}
\usepackage{opex3}

\usepackage{here}

\newcommand\pictc[5]{\begin{figure}
                   \centerline{\vspace{-0mm}
\includegraphics[width=#1\columnwidth,height=0.7\textheight,keepaspectratio]{#3}}
               \protect\caption{\protect\label{fig:#4} #5}
                \end{figure}            }

\newcommand\pict[4][1]{\pictc{#1}{!htb}{#2}{#3}{#4}}
\newcommand\rpict[1]{\ref{fig:#1}}

\newcommand\leqt[1]{\protect\label{eq:#1}}
\newcommand\reqtn[1]{\ref{eq:#1}}
\newcommand\reqt[1]{(\reqtn{#1})}

\newcommand\lsect[1]{\protect\label{sect:#1}}

\usepackage{amsmath}
\newcommand{\euler}[1]{{\usefont{U}{eur}{m}{n}#1}}
\newcommand{\umu}{\mbox{\euler{\char22}}}

\newcommand{\umum}{\mbox{\euler{\char22}}\textrm{m}}

\begin{document}

\title{Nonlinear Bloch modes in two-dimensional photonic lattices}

\author{Denis Tr\"ager$^{1,2}$, Robert Fischer$^{1}$, Dragomir N. Neshev$^{1}$,\\
Andrey A. Sukhorukov$^{1}$, Cornelia Denz$^{2}$, Wieslaw Krolikowski$^{1}$,\\
and Yuri S. Kivshar$^{1}$}

\address{$^1$Nonlinear Physics Centre and Laser Physics Centre, Centre
  for Ultrahigh bandwidth Devices for Optical Systems (CUDOS),
  Research School of Physical Sciences and Engineering, Australian
  National University, Canberra, ACT 0200, Australia\\ 
  $^2$Institut f\"ur Angewandte Physik, Westf\"alische
  Wilhelms-Universit\"at, 48149 M\"unster, Germany} 

\email{dtraeger@uni-muenster.de}
\homepage{http://www.rsphysse.anu.edu.au/nonlinear - http://www.uni-muenster.de/physik/ap/denz}

\begin{abstract}
We generate experimentally different types of two-dimensional Bloch waves
of a square photonic lattice by employing the phase imprinting technique.
We probe the local dispersion of the Bloch modes in the photonic lattice by
analyzing the linear diffraction of beams associated with the high-symmetry
points of the Brillouin zone, and also distinguish the regimes of
normal, anomalous, and anisotropic diffraction through observations of
nonlinear self-action effects.
\end{abstract}


\section{Introduction}

The study of the wave propagation in optical periodic structures
such as photonic crystals~\cite{Joannopoulos:1995:PhotonicCrystals} has attracted growing interest
in recent years. The periodic photonic structures exhibit unique
properties allowing to manipulate the flow of light at the
wavelength scale and create the basis for novel types of integrated
optical devices. In such periodic dielectric structures, the
propagation of light is governed by the familiar Bloch theorem due
to the interplay between the light and the surrounding periodic
structure~\cite{Russell:1991-1599:JMO}, that introduces the spatially extended
linear waves, the so-called  Bloch waves, as the eigenmodes of the
corresponding periodic potential. Thus, the properties of
electromagnetic waves in periodic structures are fully determined by
the Bloch wave dispersion which, for the spatial beam propagation,
represents the relation between the longitudinal and transverse
components of the Bloch wavevector. Since any finite beam
can be expressed as a superposition of such Bloch
waves~\cite{Russell:1995-585:ConfinedElectrons},  the beam
propagation in any periodic structure is also determined from the
local dispersion. In particular, the beam  propagation direction is
defined by the normal to the dispersion curve while the beam
spreading is governed by the curvature of this curve.

The study of Bloch waves and their temporal and spatial dispersion
provides a key information about overall properties of any periodic
structure. In particular, depending on the local dispersion and a
local value of the wave vector, an optical beam (or pulse)
experience normal, anomalous or even vanishing diffraction (or
dispersion)~\cite{Russell:1995-585:ConfinedElectrons,
Eisenberg:2000-1863:PRL, Pertsch:2002-93901:PRL}. 
Experimentally, the Bloch-wave character of electromagnetic waves in
photonic crystal waveguides has been deduced indirectly by detecting the
out-of-plane leakage of light~\cite{Loncar:2002-1689:APL}, by investigating the
evanescent field coupling between a tapered optical fiber and a
photonic crystal waveguide~\cite{Barclay:2004-4:APL}, and more directly by
local near-field probing of the intensity distribution in a
waveguide~\cite{Bozhevolnyi:2002-235204:PRB}. The full band structure of a photonic
crystal waveguide has been recovered very recently by employing a
near-field optical microscope and probing both the local phase and
amplitude of the light propagating through a single-line defect
waveguide~\cite{Gersen:2005-123901:PRL, Engelen:2005-4457:OE}.

The Bloch-wave dynamics in periodic structures becomes even more
dramatic in the presence of the nonlinear medium response that may
lead to the formation of strongly localized structures, discrete and
gap solitons~\cite{Voloshchenko:1981-902:ZTF, Chen:1987-160:PRL,
Christodoulides:1988-794:OL, deSterke:1994-203:ProgressOptics,
Feng:1993-1302:OL, Nabiev:1993-1612:OL, John:1993-1168:PRL,
Akozbek:1998-2287:PRE, Mingaleev:2001-5474:PRL,
Eggleton:1997-2980:JOSB}. The properties of the Bloch modes of
nonlinear periodic structures have been extensively studied in the
one-dimensional geometries, including the Bragg gratings and
waveguide arrays~\cite{Mandelik:2003-53902:PRL,
Sukhorukov:2004-93901:PRL}, as well as the study of modulational
instability of one-dimensional waves~\cite{deSterke:1998-2660:JOSB,
Meier:2004-163902:PRL, Iwanow:2005-7794:OE,
Wirth:2005-TuB2:ProcNLGW, Chavet:2005-WD39:ProcNLGW,
Eggleton:1998-267:OC}. 
More recently, the Brillouin zone structure of nonlinear two-dimensional photonic lattices was characterized based on the features of collective wave dynamics for partially coherent multi-band excitations~\cite{Bartal:2005-163902:PRL}.
Nevertheless, to the best of our knowledge, no experimental studies of individual two-dimensional Bloch waves from different bands and probing the Bloch wave local dispersion have been reported yet.

The aim of this paper is twofold. First, we probe the local spatial
dispersion of the Bloch modes of a two-dimensional optically-induced
photonic lattice by analyzing the evolution of linear and nonlinear
propagation modes associated with the high-symmetry points of the
first Brillouin zone. In particular, we excite the Bloch waves
associated with the high-symmetry points of the two-dimensional
lattice by matching their unique phase structure and observe
different regimes of the linear diffraction. Second, we employ a
strong self-focusing nonlinearity and study nonlinear self-action effects for the two-dimensional Bloch waves. This allows us to probe and characterize the spatial diffraction of each particular Bloch mode, depending on the curvature of the dispersion surfaces at the corresponding point of the Brillouin zone.

\pict[1]{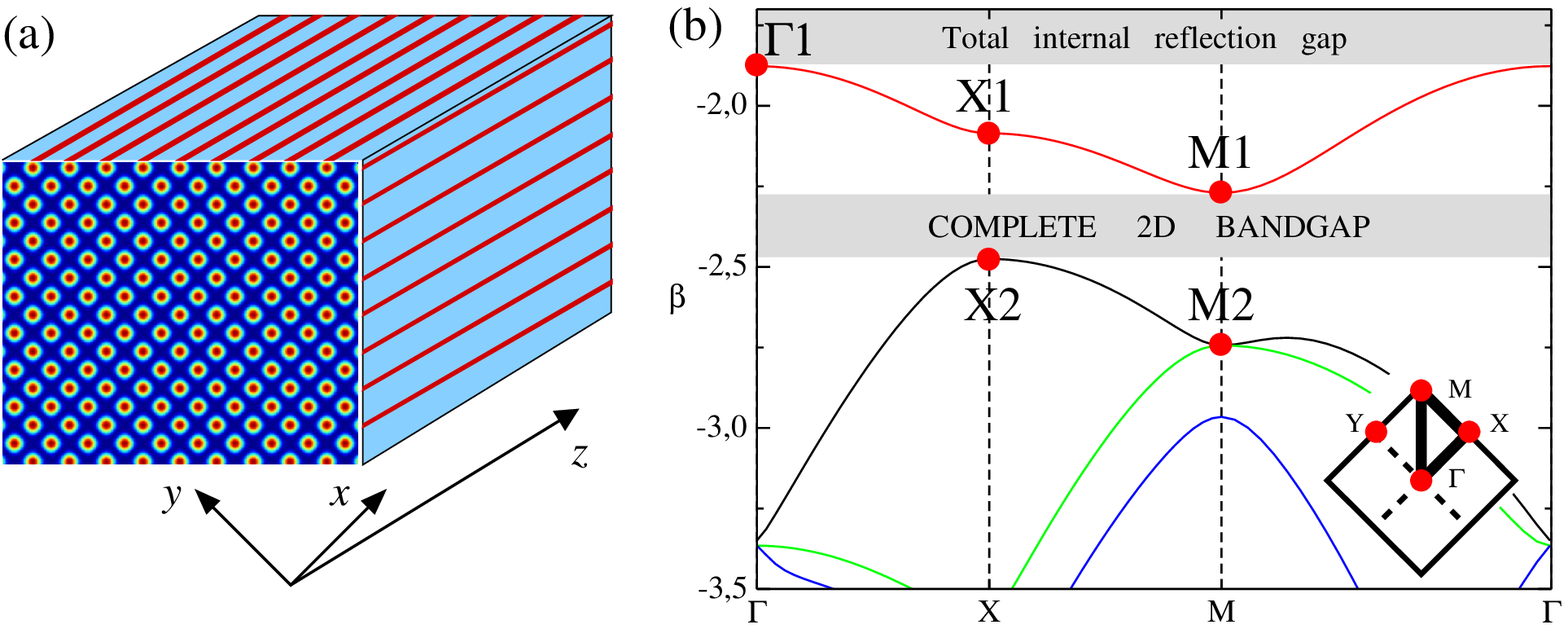}{bandgap}{(a) Experimental image of a
two-dimensional optically-induced photonic lattice, that is
spatially periodic in the transverse directions ($x,y$) and
stationary along the longitudinal direction $z$. (b) Calculated
bandgap dispersion $\beta(K)$. Dots indicate the main symmetry
points. Inset in (b) depicts the corresponding first Brillouin
zone.}

\section{Two-dimensional Bloch waves: theoretical background}
\lsect{theory}

We study the propagation of an extraordinary polarized optical
beam (a probe beam) in a biased photorefractive crystal with an
optically induced two-dimensional photonic lattice. We consider 
a spatially periodic pattern of the refractive index in the form of a  square lattice, which is stationary in the longitudinal ($z$)
direction [Fig.~\rpict{bandgap}(a)]. The photonic lattice is formed
by the interference of four mutually coherent ordinary polarized
optical beams. This interference pattern [Fig.~\rpict{bandgap}(a)],
\[
 I_p(x,y)=I_g \{\cos[\pi (x+y)/d]+\cos[\pi (x-y)/d]\}^2 ,
\] 
induces a refractive index modulation of the crystal  for extraordinary
polarized light via the strong electro-optic
effect~\cite{Efremidis:2002-46602:PRE}. Here $x$ and $y$ are the
transverse coordinates, and $d$ is the lattice period. The spatial
evolution of the extraordinary polarized beam with a slowly varying
amplitude $E(x,y,z)$ propagating along the lattice is then governed
by the following nonlinear parabolic equation,
\begin{equation} \leqt{nls}
   i \frac{\partial E}{\partial z}
   + D \left(   \frac{\partial^2 E}{\partial x^2}
              + \frac{\partial^2 E}{\partial y^2} \right)
   + {\cal F}(x,y, |E|^2) E = 0,
\end{equation}
where
\begin{equation} \leqt{field}
  {\cal F}( x, y, |E|^2) = - \frac{\gamma }{I_b + I_p(x,y) + |E|^2}
\end{equation}
describes the refractive index change that includes the
two-dimensional lattice itself and the self-induced index change
from the probe beam. The parameters used for numerical calculations
are chosen to match the conditions of typical experiments discussed
below: the dimensionless variables $x$, $y$, $z$  are normalized to
the typical scale 
$x_s=y_s=1~\umu{}\mathrm{m}$, and $z_s=1\,$mm, respectively; the
diffraction coefficient is $D = z_s \lambda / (4 \pi n_0 x_s^2)$; $n_0
= 2.35$ is the refractive index of a bulk photorefractive crystal, 
$\lambda=532\,$nm is the laser wavelength in vacuum, the parameter
$I_b=1$ is the constant dark irradiance, $\gamma = 2.36$ is a
nonlinear coefficient proportional to the electro-optic coefficient
and the applied DC electric field, lattice modulation is $I_g=0.49$,
and $d=23\,\umu$m is the lattice period. 

Such a periodic modulation of the refractive index results in the
formation of a bandgap spectrum for the transverse components of the
wave vectors $K_x$ and $K_y$. Then the propagation of linear waves
through the lattice is described by the spatially extended
two-dimensional eigenmodes, known as the two-dimensional Bloch
waves. They can be found as solutions of linearized
equation~\reqt{nls} in the form
\begin{equation} \leqt{Blochwave}
  E(x,y;z) = \psi(x,y) \exp( i \beta z + i K_x x + i K_y y),
\end{equation}
where $\psi(x,y)$ is a periodic function with the periodicity of the
underlying lattice, and $\beta$ is the propagation constant. For a
square lattice shown in Fig.~\rpict{bandgap}(a), the dispersion
relation $\beta( K_x, K_y )$ is invariant with respect to the
translations $K_{x,y} \rightarrow K_{x,y} \pm 2 \pi / d$, and
therefore is fully defined by its values in the first Brillouin zone
[Fig.~\rpict{bandgap}(b, inset)]. The dispersion relation
$\beta(K_x, K_y )$ for this lattice is shown in
Fig.~\rpict{bandgap}(b) where the high-symmetry points of the
lattice are marked by red dots. 

It is important to note that, for the chosen lattice parameters, there exists a full two-dimensional band gap between the first and
the second spectral band. The existence of a typical bandgap
structure of the lattice with a complete two-dimensional gap and the
highly nonlinear properties of the photorefractive crystal make the
optically-induced photonic lattice a direct {\em analog of a  
two-dimensional nonlinear photonic crystal}. Therefore, our
experiments offer an ideal test-bench for the similar phenomena with highly
nonlinear and tunable two-dimensional photonic crystals that may be studied in the future with fabricated structures in nonlinear materials.

\pict{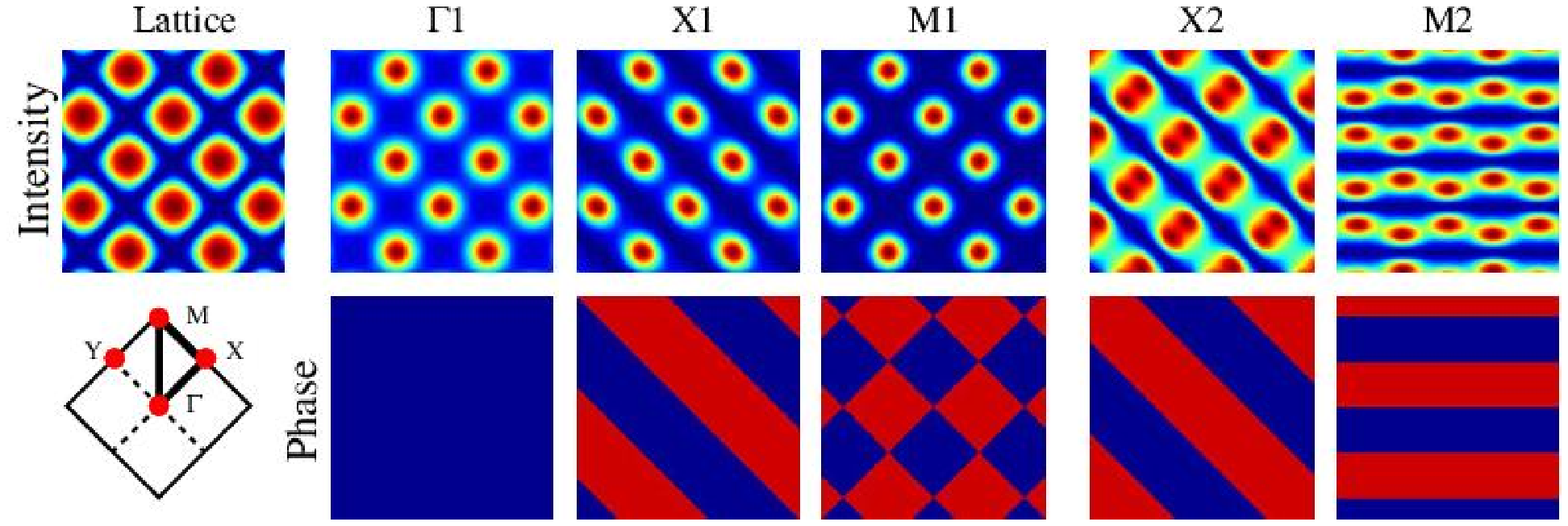}{blochmodes}{Intensity (top) and phase
(bottom) of different Bloch modes from the high symmetry points of
the first and second band of a square lattice. The blue color for
the phase distribution corresponds to the zero phase, while the red
color corresponds to the $\pi$ phase. }

The intensity and phase structure of the calculated Bloch waves
for the high symmetry points of the lattice from the first and
second spectral bands are shown in Fig.~\rpict{blochmodes}. The
upper row shows the Bloch-wave intensity profiles and the bottom
row shows the corresponding phase structure. As a reference, the
first column shows the light intensity of the lattice itself. For
the two-dimensional Bloch waves from the first band, the intensity
distribution of all modes is reflecting the structure of the square
lattice, with the intensity maxima coinciding with those of the
lattice. However, the phase structure differs substantially. As can
be seen from Fig.~\rpict{blochmodes}, the phase of the
two-dimensional Bloch waves originating from the $\Gamma_1$ point is
constant. The phase structure becomes nontrivial for the modes from
the X$_1$ and M$_1$ points. For the X$_1$ (Y$_1$) point, the phase represents a stripe-like pattern being constant along one principal direction of the lattice and exhibiting $\pi$ phase jumps along the other
direction. For the Bloch waves originating from the M$_1$ point the
phase distribution resembles a chessboard pattern.

On the other hand, the two-dimensional Bloch modes from the second
spectral band have the intensity maxima centered between the maxima
of the square lattice. The phase structure has a form of stripes
oriented along one of the principal directions of the
two-dimensional lattice for the X$_2$ point, or in $45^\circ$ with
respect to the principal axes for the M$_2$ point. The $\Gamma_2$ point
appears to be nearly degenerate with the propagation constants
nearly the same for the second, third, and fourth bands, and we do
not consider it here.

The difference in the phase structure of the two-dimensional Bloch
waves translates into the differences in propagation dynamics for beams of a finite size which spectrum is localized in the vicinity of the corresponding high-symmetry points in the Brillouin zone.
Indeed, the alternating phase is a signature of strong Bragg scattering, that may lead to an enhanced diffraction of beams similar to the
effect of the dispersion enhancement in the Bragg
gratings~\cite{Slusher:2003:NonlinearPhotonic}. Therefore, the beams can experience anisotropic diffraction due to the asymmetric phase structure of the corresponding Bloch waves, and this can be detected by analyzing the beam broadening in the linear regime.

The sign of the curvature of the related dispersion surface can be
identified experimentally utilizing the nonlinear self-action of the
beam. In the case of a medium with positive (self-focusing)
nonlinearity, increasing input beam intensity will result in either
focusing or defocusing of the output beam depending on whether the
curvature of the dispersion surface is convex or concave,
respectively. A close examination of the spatial dispersion defined
by the bandgap spectrum of the lattice [Fig.~\rpict{bandgap}(b)]
shows that the beams associated with the $\Gamma_1$ and X$_2$ points
will experience self-focusing in both $(x,y)$ directions due to the
convex curvature along the $x$ and $y$ directions. On the other
hand, the beams associated with the M$_1$ point will experience
nonlinear self-spreading due to the concave curvature at the
corresponding point of the dispersion surface. Totally different
behavior is expected for the beams associated with the X$_1$ point of
the lattice spectrum, as the curvatures of the dispersion surface
are opposite in the $x$ and $y$ directions. Such beams will
experience an anisotropic nonlinear response: they will focus along
the direction of the constant phase and at the same time will
self-defocus in the orthogonal direction. The symmetry point M$_2$ of
the dispersion curve is degenerate between the second and the third
bands, with both bands having opposite but isotropic curvatures. Due
to this degeneracy the nonlinear self-action of the beams associated
with this point will result in a complex beam dynamics.

\section{Experimental arrangements} \lsect{exp}

\pict{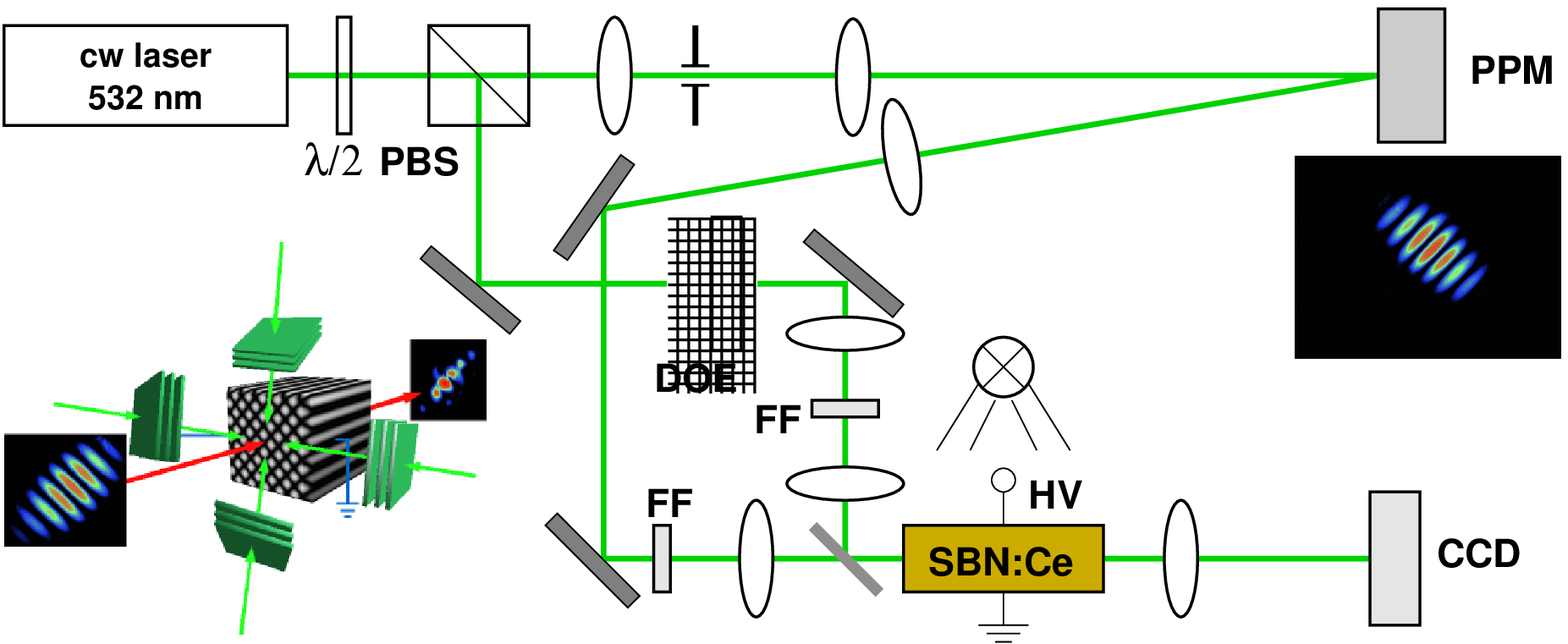}{setup}{Experimental setup for the excitation of
two-dimensional Bloch modes: HV: High voltage, CCD: camera, FF:
Fourier filter mask, $\lambda/2$: half wave plate; PPM: Programmable
phase modulator, DOE: Diffractive optical element to produce four
coherent beams, PBS: Polarizing beam splitter. Left inset: Geometry
of the two-dimensional optical lattice. Right inset: Example of a
phase and amplitude engineered wave in the optical lattice.}

In order to study experimentally the generation, formation and
propagation of linear and nonlinear Bloch waves in two-dimensional
photonic lattices, we implement the setup shown schematically in
Fig.~\rpict{setup}. An optical beam from a cw frequency doubled
Nd:YVO$_4$ laser, at the wavelength of 532\,nm, is split by a
polarizing beam splitter into two beams with orthogonal
polarizations. The vertically polarized beam is passed through a
diffractive optical element (DOE), which produces two orthogonally
oriented pairs of beams. An optical telescope combines these four
beams at the input face of the photorefractive crystal, thus
forming a two-dimensional square interference pattern which is
stationary along the crystal length (see inset in
Fig.~\rpict{setup}). The period of this pattern is $23\,\umum$. The
crystal is a Cerium doped SBN:60 of $20\,$mm $\times$ $5\,$mm $\times$
$5\,$mm  biased externally with a DC electric field of 4\,kV/cm
applied along the c-axis (horizontal). Due to a strong anisotropy of
the electro-optic effect, the ordinary polarized lattice beams will
propagate linearly inside the crystal, while in the same time
inducing a refractive index modulation for the extraordinary
polarized (probe) beam~\cite{Efremidis:2002-46602:PRE}.

The second, extraordinary polarized laser beam is expanded by a
telescope and illuminates the active area of a Hamamatsu
programmable phase modulator (PPM). The modulated beam is then
imaged by a large numerical aperture telescope at the input face of
the photorefractive crystal. A spatial Fourier filter (FF) is placed in the
focal plane of the telescope to eliminate higher-order spectral
components and ensure that the optical beam entering the crystal will
have the phase and amplitude structure required to match the specific
Bloch mode. The modulated probe beam is combined with the lattice onto
a beam splitter. Thus it will propagate onto the induced periodic
index modulation and simultaneously will experience a strong nonlinear
self-action at sub-micro-Watt range, due to the strong photorefractive 
nonlinearity. Both faces of the crystal can be imaged on a CCD
camera by a high numerical aperture lens to capture beam intensity
distribution.

In order to excite selectively different Bloch modes of the
two-dimensional lattice, the optical beam must match
their transverse amplitude structure. This is achieved by the use of
PPM that converts the initially Gaussian probe beam into the desired
amplitude and phase modulation at the front face of the
photorefractive crystal. For low input intensity the incident probe
beam, representing linear Bloch wave, does not affect the refractive
index of the lattice and hence its propagation is completely
determined by the dispersion at the particular point of the
Brillouin zone. A finite beam will diffract with a rate depending on
the value of the diffraction coefficients along the principal
directions of the lattice. These diffraction coefficients are
determined by the curvature of the dispersion surfaces along the $x$
and/or $y$ directions. With increasing power, nonlinear self-action
of the beam will counteract its diffraction in the case of normal
diffraction, but it will enhance the beam spreading in the case of
the anomalous diffraction. These features of the nonlinear self-action of finite beams allows us to identify the character of the dispersion curves when the beam is associated with a specific Bloch mode of the lattice.

\pict{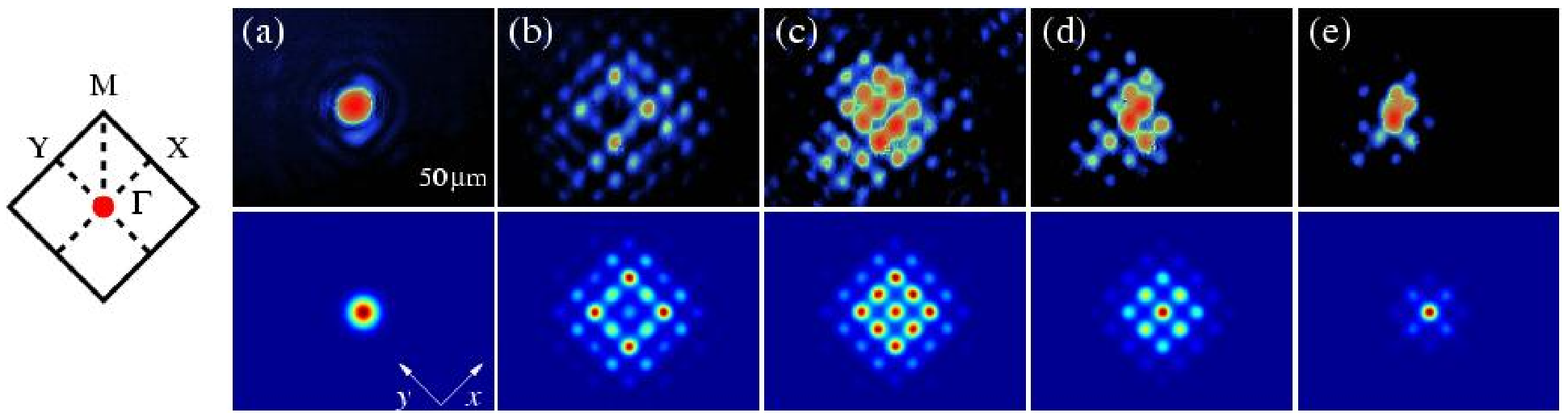}{G1}{Experimental data (top row) and numerical
results (bottom) for the excitation of the two-dimensional Bloch
waves from the $\Gamma$ point of the first spectral band
($\Gamma_1$). (a) Input beam; (b-e) outputs for input powers of
25\,nW, 125\,nW, 250\,nW, and 375\,nW, respectively.}

Our experiments are complemented by the numerical simulations of the
underlying equation~\reqt{nls} with the initial conditions matching
the transverse structure of the corresponding Bloch wave
superimposed on a Gaussian carrier beam
\begin{equation} \leqt{Gauss}
  E(x,y) = A \exp(x^2/w_x^2 + y^2/w_y^2),
\end{equation}
where $A$ is a constant amplitude, $w_x$ and $w_y$ are the
corresponding beam widths along the $x$ and $y$ axes, respectively.
Our numerical simulations allow us to trace, with a high accuracy, the
actual beam evolution inside the crystal that is not directly accessible in
experiment, as well as provide the opportunity to test the beam
evolution  for larger propagation distances beyond the
experimentally accessible crystal lengths.

\section{Excitation of the Bloch modes of the first band}

First, we study experimentally the propagation of beams associated
with different Bloch waves from the first spectral band of the
lattice bandgap spectrum (Fig.~\rpict{bandgap}).

\subsection{$\Gamma_1$-point}

The excitation of the point $\Gamma_1$ is realized simply by launching
a Gaussian beam [Eq.~(\ref{eq:Gauss})] along the lattice and having
zero transverse wavevector components. The structure of this Bloch
wave is fully symmetric along the principal axes of the lattice
(Fig.~\rpict{blochmodes}). If the initial beam excites a single
lattice site, then the diffraction output represents a typical
discrete diffraction~\cite{Fleischer:2003-147:NAT,
Martin:2004-123902:PRL} and it is well suited to characterizing the
induced periodic potential. When the intensity of the initial beam
is high enough, the nonlinearity induced index change leads to a shift
of the propagation constant inside the total internal reflection gap
[Fig.~\rpict{bandgap}(b)] and gives rise to the formation of discrete
lattice solitons~\cite{Fleischer:2003-147:NAT,
Martin:2004-123902:PRL}.

Our experimental results were performed with an input beam of width
$w_x=w_y=18\,\umum$. For low input powers of $25\,$nW [see
Fig.~\rpict{G1}(a)] the beam undergoes strong discrete diffraction on
the lattice, where most of its energy is transferred to the outside
lobes. With increasing the laser power [Fig.~\rpict{G1}(c-e)] the beam
self-focuses leading to the formation of a discrete lattice soliton in
agreement with previous experimental
studies~\cite{Fleischer:2003-147:NAT, Martin:2004-123902:PRL}. The  
numerical simulations shown in Fig.~\rpict{G1} (bottom row) are in
good agreement with the experimental observations.

\subsection{X$_1$-point}

\pict[0.65]{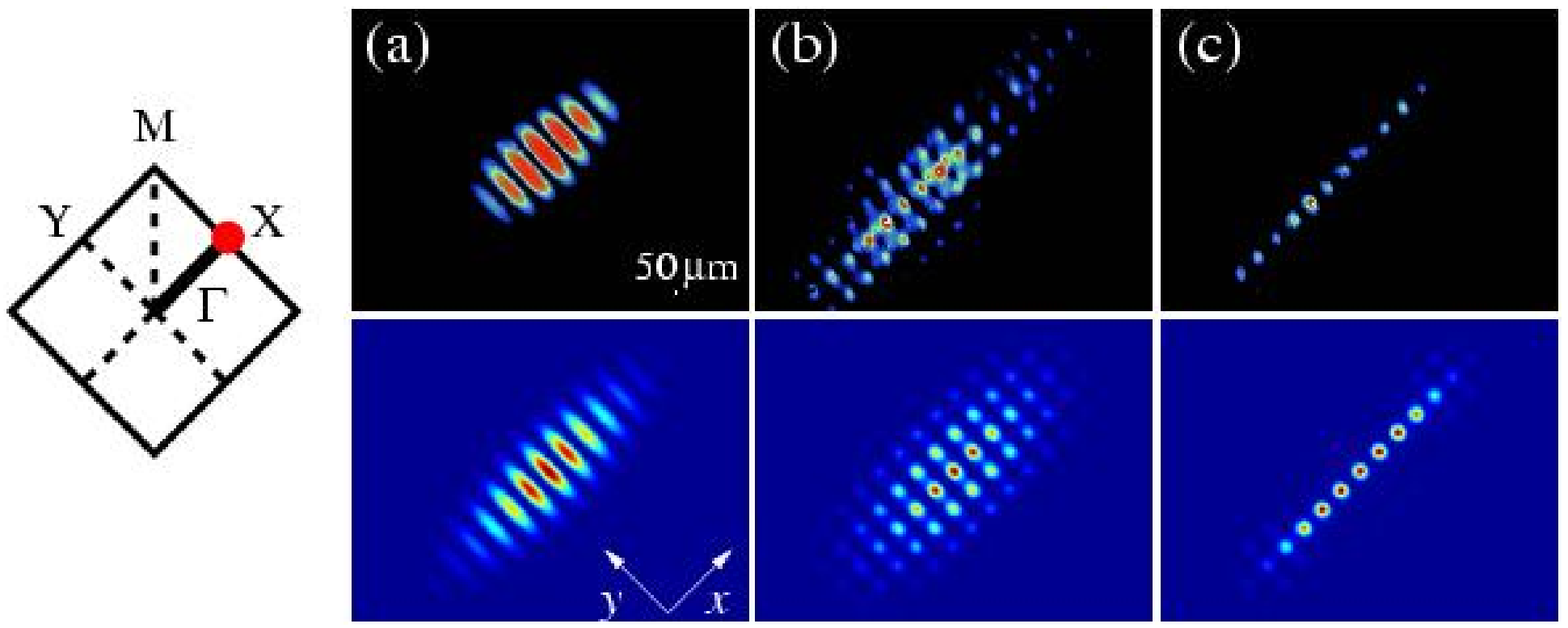}{X1}{Experimental data (top row) and numerical
results (bottom) for the excitation of the two-dimensional Bloch
waves from the X point of the first spectral band (X$_1$). (a) Input
beam; (b,c) outputs for input powers of 25\,nW and 375\,nW,
respectively.} 

The Bloch wave at the X symmetry point of the first band has a
strongly asymmetric phase structure. This leads to
anisotropic diffraction for the propagating beams
associated with this Bloch mode, allowing for new types of
waveguiding~\cite{Cohen:2004-103902:PRL, Freedman:2005-1349:JOSB} due
to different curvatures of the dispersion surface in $x$ and $y$
directions. 
In order to balance the rate of beam broadening due to diffraction
along these directions, in experiment the input beam is made
elliptical, elongated along the $x$ axis. 
Its phase is constant along the $y$ direction and alternates by $\pi$
along the perpendicular $x$ direction [Fig.~\rpict{X1}(a)]. 
This stripe-structure is launched on site, i.e. with position of the
intensity maxima on lattice sites. In numerics, the input profile is
modeled by the following expression
\[
   E(x,y) = A \cos(K x)\exp(x^2/w_x^2 + y^2/w_y^2),
\]   
where $w_x=100\,\umum$, $w_y=33\,\umum$ and $K=\pi/d$ is the lattice
wavevector. 

Our experimental results and the corresponding numerical simulations
show the same behavior for the nonlinear response of the beam at the
output face of the crystal [Fig.~\rpict{X1}(b,c) top and bottom row
respectively]. At low laser powers, the initial beam spreads
strongly in $x$-direction due to the larger curvature of the
dispersion surface. Increasing beam power leads to strong focusing
of the beam along $y$ direction and beam spreading along $x$ axis. This
difference in the nonlinear self-action of the beam allows one to
identify experimentally that the dispersion surface has opposite
curvatures in two principal directions of the lattice as follows from the theoretically calculated band-gap diagram [Fig.~\rpict{bandgap}(b)].
The process of strong focusing along the non-modulated $y$ direction
is closely related the effect of grating mediated
waveguiding~\cite{Cohen:2004-103902:PRL, Freedman:2005-1349:JOSB}.

\subsection{M$_1$-point}

\pict{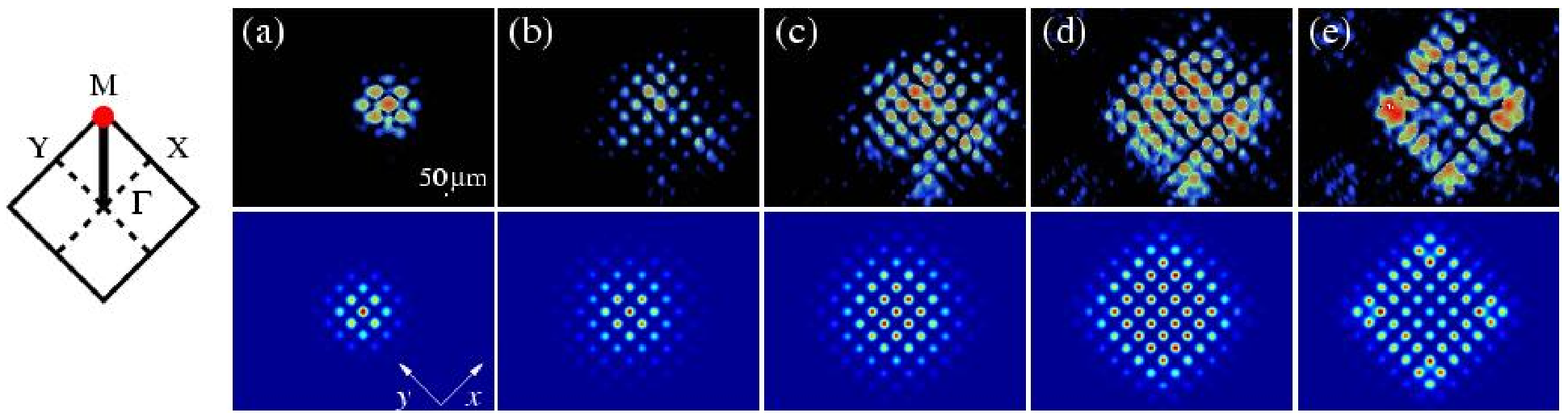}{M1}{Experimental data (top row) and numerical results
  (bottom) for the excitation of the two-dimensional Bloch waves from
  the M point of the first spectral band (M$_1$). (a) Input beam;
  (b-e) outputs for input powers of $40\,$nW, $125\,$nW, $300\,$nW,
  and $850\,$nW, respectively.} 

The structure of the dispersion surface near the M symmetry point of
the first band is symmetric in $x$ and $y$ directions. To match the
Bloch-wave profile, the input beam is modulated such that it
represents humps of alternating phase in the form 
\[
    E(x,y)= A\cos(K x)\cos(K y)\exp(x^2/w_x^2 + y^2/w_y^2),
\]
with $w_x=w_y=51\,\umum$ [Fig.~\rpict{M1}(a)].

The curvature of the dispersion surface is concave as indicated in
Fig.~\rpict{bandgap}(b). Therefore the initial beam is expected to
exhibit enhanced defocusing with increasing of the beam power. Our
experimental measurements of the output beam intensity distribution
are depicted in Fig.~\rpict{M1}. At low laser powers ($P=40\,$nW) the
beam diffracts linearly forming a Bloch state from the M$_1$-symmetry 
point, shown in Fig.~\rpict{M1}(b). With increasing power
[Fig.~\rpict{M1}(c-e) at power levels $125\,$nW, $300\,$nW, and
$850\,$nW, respectively] the beam defocuses as expected and forms a
square type pattern [Fig.~\rpict{M1}(e)]. Similar behavior is also
observed in the performed numerical simulations [Fig.~\rpict{M1},
bottom row].

\section{Excitation of the Bloch modes of the second band}

The second band of the lattice bandgap spectrum is separated from
the first band by a two-dimensional photonic gap. The Bloch modes
from the top of the second band (as the X symmetry point) then can
be moved by nonlinearity inside the gap, leading to the formation of
spatially localized gap solitons. On the other hand, the second band
overlaps with the higher-order bands at the $\Gamma$ and M points
leading to degeneracy of the two-dimensional Bloch modes and
subsequently complex beam dynamics that are reproducible in
numerical simulations and experiments but difficult to interpret. Out of these degenerate points, below we consider only the M symmetry point.

\subsection{X$_2$-point}

In order to match the profile of the Bloch wave from the X symmetry
point of the second band we use a modulated Gaussian beam, of the
form
\[
    E(x,y)=A\cos[K(x-d/2)]\exp(x^2/w_x^2 + y^2/w_y^2),
\]
where the maxima of this modulated pattern [Fig.~\rpict{X2}(a)] are
shifted with respect to the lattice maxima by half a lattice period
along the $x$ axis. The structure of the dispersion surface of this
Bloch mode is highly anisotropic, therefore the beam diffracts
differently in $x$ and $y$ directions. To account for this anisotropic
diffraction we used an elliptic beam elongated along the 
$x$ axis with $w_x=100\,\umum$ and $w_y=33\,\umum$. At low
laser powers ($20\,$nW) the beam diffracts linearly, while
reproducing the structure of the Bloch wave from the X point of the
second band (Fig.~\rpict{blochmodes}). With increase of the laser
power [$50\,$nW, $100\,$nW, and $200\,$nW for Fig.~\rpict{X2}(c-e),
respectively] the beam focuses in both transverse directions and
forms a strongly localized state [Fig.~\rpict{X2}(e)]. Such state represents the theoretically predicted gap solitons in photonic
crystals~\cite{John:1993-1168:PRL, Akozbek:1998-2287:PRE}. It has a
reduced symmetry with respect to the lattice and it is formed by the
combined action of Bragg reflection in $x$-direction and total
internal reflection in
$y$-direction~\cite{Fischer:physics/0509255:ARXIV}. The experimental 
data are in excellent agreement with the numerical simulations
[Fig.~\rpict{X2}(bottom row)].

We note that a symmetric superposition of X$_2$ and Y$_2$
states gives rise to symmetric gap
solitons~\cite{John:1993-1168:PRL, Akozbek:1998-2287:PRE} or gap
vortices~\cite{Manela:2004-2049:OL, Bartal:2005-53904:PRL}.

\pict{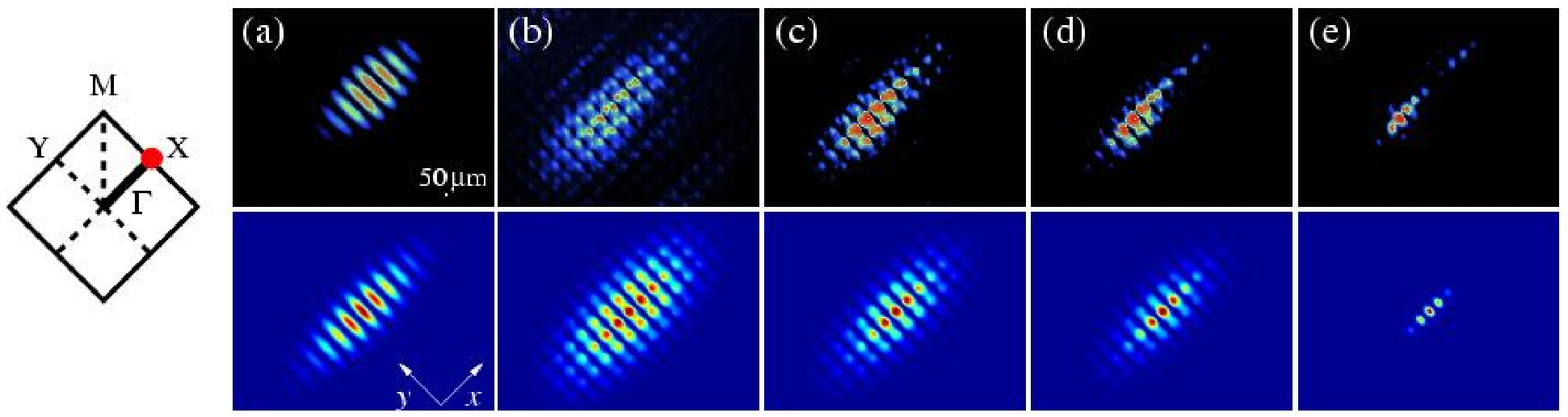}{X2}{Experimental data (top row) and numerical results
  (bottom) for the excitation of the two-dimensional Bloch waves
  from the X point of the second spectral band (X$_2$). (a) Input
  beam; (b-e) outputs for input powers of $20\,$nW, $50\,$nW,
  $100\,$nW, and $200\,$nW, respectively.}

\subsection{M$_2$-point}

\pict{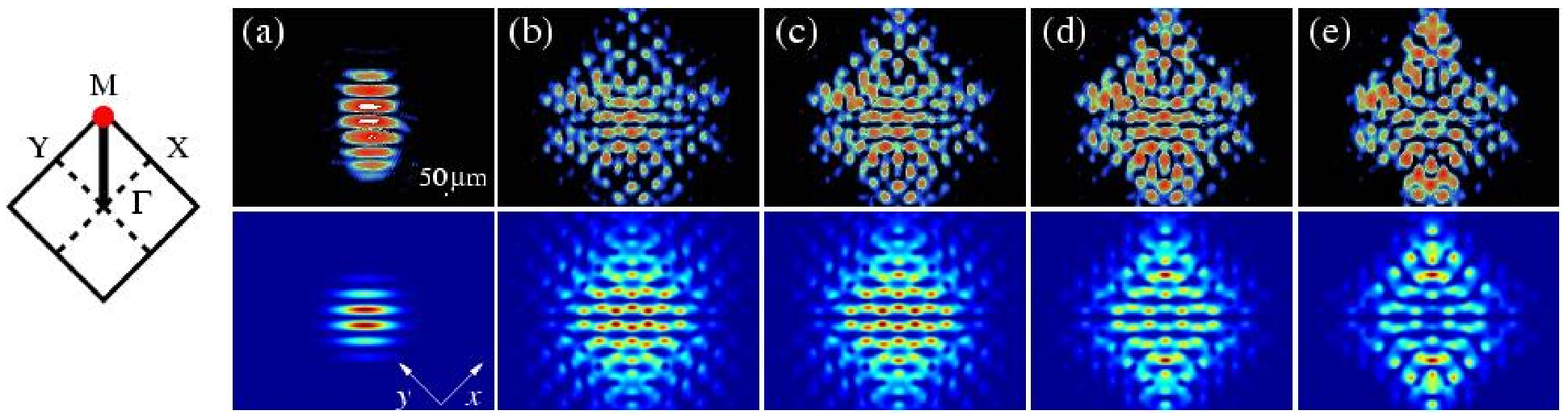}{M2}{Experimental data (top row) and numerical
results (bottom) for the excitation of the two-dimensional Bloch
waves from the M point of the second spectral band (M$_2$). (a) Input
beam; (b-e) outputs for input powers of $25\,$nW, $90\,$nW, $270\,$nW,
and $660\,$nW, respectively.}

The M symmetry point of the second band is degenerate as the
propagation constant coincides with that of the third band.
Furthermore, two dispersion curves have opposite curvatures and
therefore a complex beam dynamics is expected. 
We select the Bloch mode [Fig.~\rpict{blochmodes}] which structure can
be approximated by horizontal stripes, that are oriented at a 
45$^\circ$ angle with respect to the principal axes of the lattice.
To match this Bloch mode we used a Gaussian beam
modulated at 45$^\circ$ with respect to the $x$ and $y$ axes
[Fig.~\rpict{M2}(a)],
\[
    E(x,y) = A\cos(K x + K y)\exp(x^2/w_x^2 + y^2/w_y^2),
\]
with $w_x=w_y=38\,\umum$.

Linear propagation at low power is shown in Fig.~\rpict{M2}(b),
where the beam is strongly diffracting at the crystal output. At
higher powers in (c) $90\,$nW, (d) $270\,$nW and (e) $660\,$nW the nonlinear
self-action leads to strong beam reshaping. The central part of the
beam experiences self-defocusing while the intensity in the outer
region increases. Again, our experimental results are in good
agreement with the numerical simulations [Fig.~\rpict{M2}, bottom
row].

\section{Conclusions}

We have generated experimentally and analyzed theoretically different
types of two-dimensional Bloch waves of a square photonic lattice by
employing the phase imprinting technique. We have excited selectively
the Bloch waves belonging to different high-symmetry points of the
two-dimensional photonic lattice, and demonstrated the unique linear
and nonlinear anisotropic properties of the lattice dispersion
resulting from the different curvatures of the dispersion surfaces of
the first and second spectral bands. We have employed strong
self-focusing nonlinearity of biased photorefractive crystals to
study, for the first time to our knowledge, nonlinear self-action
effects for the two-dimensional Bloch waves at the high-symmetry
points of both fundamental and higher-order spectral bands. We have
demonstrated that our experimental results are in an excellent
agreement with the numerical simulations of both linear and nonlinear
effects of the light propagation.

\section*{Acknowledgments}

The authors acknowledge a support of the Australian Research Council,
and thank Mark Saffman for providing a photorefractive
crystal. D.T. thanks Nonlinear Physics Centre of the Research School
of Physical Sciences and Engineering for a warm hospitality during his
stay in Canberra, and also acknowledges grant and a travel support
from Konrad-Adenauer-Stiftung e.V. 

\end{document}